\documentclass[tbtags, conference]{IEEEtran}
\IEEEoverridecommandlockouts
\usepackage{epic}
\usepackage{eepic}
\usepackage[cmex10]{amsmath}
\usepackage{amssymb}
\usepackage{mathrsfs}			
\usepackage{amsthm}
\usepackage{hhline}
\usepackage[dvips]{color}
\usepackage{threeparttable}
\usepackage{stfloats}
\newtheorem{proposition}{Proposition}

\newcommand{\xn}{\boldsymbol{x}}    	
\newcommand{\zn}{\boldsymbol{z}}        
\newcommand{\pn}{\ensuremath{\boldsymbol{p}}}	
\newcommand{\Sdkr}{\mathscr{S}}		
\newcommand{\Sdklr}{\hat{\Sdkr}}	
\newcommand{\suchthat}{\;\colon\;}	
\newcommand{\CCS}{C}
\newcommand{\Fz}{z^*}
\newcommand{\phase}{\theta_m}
\newcommand{\FSS}{C}
\newcommand{\FSSl}{\widehat C}
\newcommand{\trig}{\sigma}
\newcommand{\trigR}{\tilde \trig}
\newcommand{\prou}{\omega_n}		
\newcommand{\nr}{\tilde n}
\newcommand{\chrg}{{\sigma}}
\newcommand{\trigntm}{\trig_{ \nr , m }}
\newcommand{\trigntjm}{\trig_{ \nr - j , m }}
\newcommand{\trigntmR}{\tilde \trig_{ \nr , m }}
\newcommand{\trigntmL}{\hat \trig_{ \nr - j , m }}
\newcommand{\rdm}[1]{\rho_{ #1 , m }}
\newcommand{\ann}{\mathscr{A}_{ \chrg_m , \rdm1 , \rdm2 }}
\newcommand{\annn}{\mathscr{A}_{ \trigntm , \rdm1 , \rdm2 }}
\newcommand{\annj}[1]{\mathscr{A}_{ \trig_{ \nr , #1 } , %
                                    \rho_{ 1 , #1 } , \rho_{ 2 , #1 } }}
\newcommand{\annS}{\boldsymbol{A}_{ \boldsymbol{\trig}_{\nr} , %
                                    \boldsymbol{\rho}_1 , %
                                    \boldsymbol{\rho}_2 }}
\newcommand{\annnj}{\mathscr{A}_{ \trig_{ \nr - j , m } , \rdm1 , \rdm2 }}
\newcommand{\inring}{\delta}
\title{An Enumerative Method for Encoding Spectrum Shaped %
       Binary Run-Length Constrained Sequences}
\author{Oleg~F.~Kurmaev
 \thanks{The author is with the Moscow Institute of
         Electronic Engineering (MIEE/MIET),
         124498, Moscow, Russia.
         (e-mail: {\tt \symbol{"3C}kurmaev@org.miet.ru\symbol{"3E}})}}
\date{}
\begin{document}
 \maketitle
 \begin{abstract}
  A method for encoding and decoding
  spectrum shaped binary run-length constrained sequences
  is described.
  The binary sequences with predefined range
  of exponential sums are introduced.
  On the base of Cover's enumerative scheme,
  recurrence relations for calculating the number
  of these sequences are derived.
  Implementation of encoding and decoding procedures
  is also shown.
 \end{abstract}

 \section{Introduction}

 Binary sequences with constrained run length of zeros
 are known in literature as $dk$ sequences~\cite{Immink2004, ImminkSiegelWolf98}.
 In these sequences, single ones are separated by at least $d$,
 but not more then $k$ zeros.
 A $dkr$ sequence is a $dk$ sequence, ending in a run
 of not more then $r$ trailing zeros.
 A $dklr$ sequence is a $dkr$ sequence, beginning with a run of
 not more then $l$ leading zeros.
 Among these, sequences with spectral null at zero frequency (dc) are notable.
 This implies a line encoding technique using NRZI rules.

 By NRZI encoding~\cite{ImminkSiegelWolf98}
 we understand mapping the source binary sequence $\xn \in \{0 , 1\}^n $
 to bipolar sequence $\zn \in \{-1, 1\}^n$
 such that
 \begin{align*}
  z_j &=\begin{cases}
    z_{j-1}, & x_j=0,\\
   -z_{j-1}, & x_j=1,
  \end{cases}\\
  z_0 &=1.
 \end{align*}

 The discrete Fourier transform (DFT) of this sequence is defined as
 \[
  \Fz_m = \sum_{ j = 0 }^{ n - 1 }
           z_{ j + 1} e^{ -2 \pi i \frac{ m j }{ n } } ,
           \quad m = 0 , \dots , n - 1 .
 \]

 The usual spectrum shaping requirements are first,
 a zero-mean value for the transmitted data sequence, and second,
 small power content at low frequencies.
 This second requirement can be written as
 \[
  | \Fz_m | < M ,
 \]
 where $ M $ is some constant.
 These zero-mean value sequences are called DC-free RLL
 or DCRLL~\cite{Braun00anenumerative}.

 Although a study of spectral properties of channel sequences
 was performed by Nyquist~\cite{Nyquist28} as early as in the late 20s;
 regular papers concerning this problem were appeared
 in the late 60s~-- early 70s with the contributions coming from
 Gorog~\cite{Gorog68}, Franklin and Pierce~\cite{FranklinPierce72},
 and some other authors.
 In 1984 Pierobon~\cite{Pierobon84} proved
 that the finite running digital sum condition
 is a necessary and sufficient condition
 for zero mean and spectral density vanishing at zero frequency.
 Later, Marcus and Siegel~\cite{MarcusSiegel87}
 have expanded the concept of the running digital sum
 to each component of the DFT.
 This allows extension of the spectral null control
 from zero up to the Nyquist frequency.

 Enumerative encoding and decoding methods for DC-free sequences were suggested by
 Norris and Bloomberg~\cite{NorrisB81},
 Immink~\cite{Immink85PhD}, 
 Vasilev~\cite{Vasilev91en},
 Braun and Immink~\cite{Braun00anenumerative}.

 A method for enumerative encoding these sequences
 with predefined dc component of the DFT $ \Fz_0 =\sum_{j=1}^n z_j$,
 which is usually called a digital sum or a charge of the sequence $\zn$,
 was suggested in~\cite{Kurmaev06}.
 Observe that $ \Fz_0 \in [-n, n]$, where $ \Fz_0 $
 admits even values whenever $n$ is even and odd values whenever $n$ is odd.

 \section{Spectrum Shaping} \label{Sec:SpSh}

 Let $\{0, 1\}^n$ be the set of all binary sequences of length $n$
 and let $\xn=(x_1, x_2, \dots, x_n)$ denote a generic element of this set.
 Let $ \Sdkr(n)=\{\xn \in \{0, 1\}^n \; |$
 satisfies the $d, k,    r$ constraints and $ x_1 = 1 \}$ and
 let $\Sdklr(n)=\{\xn \in \{0, 1\}^n \; |$
 satisfies the $d, k, l, r$ constraints$\}$.

 Let $\CCS_n^\chrg$ be the number of sequences from $ \Sdkr(n) $;
 these sequences have charge $ \chrg = \Fz_0 $.
 Using Cover's method~\cite{Cover73},
 the number of these sequences can be computed, as is shown in~\cite{Kurmaev06},
 using recurrence relation
 \begin{equation}
  \label{Eq:Cc}
  \CCS_n^\chrg=\begin{cases}
                0,                          & n<d+1,\\
                \sum_{j=d+1}^{\min(n, k+1)}
                 \CCS_{n-j}^{-\chrg-j},     & d+1\leq n
               \end{cases}
 \end{equation}
 with initial conditions
 \begin{equation*}
  \CCS_n^{-n}=\begin{cases}
               1, & n\leq r+1,\\
               0, & \text{otherwise}.
              \end{cases}
 \end{equation*}

 We might expand this method to $ \Fz_1 \dots \Fz_{ n - 1 } $
 components of the DFT,
 but it most likely will not have meaning.
 Indeed, recall that the DFT is a one-to-one transformation.
 If we go from the single spectral component to the vector
 $ \boldsymbol{ \Fz } = ( \Fz_0 , \Fz_1 , \dots , \Fz_{ n - 1 } ) $,
 then for an exact value of
 $ \boldsymbol{ \chrg } = ( \chrg_0 , \chrg_1 , \dots , \chrg_{ n - 1 } ) %
   \in \mathbb C^n $
 we see that $ \CCS_n^{ \boldsymbol{ \chrg } } $,
 can be either equal to $ 1 $ whenever $ \boldsymbol{ \chrg } = \boldsymbol{ \Fz } $
 or equal to $ 0 $ otherwise.
 Therefore, we must consider the numbers $ \CCS_n^{ \chrg_m } $
 for which spectral components $ \Fz_m $ lie in some area;
 let this area be a ring $ \ann $
 centered at $ \chrg_m $ with inner radius $ \rdm1 \in \mathbb R $
 and outer radius $ \rdm2 \in \mathbb R $; i.e. the ring
 $ \ann $ is a set
 $  \ann
   = \{ \chrg_m
      \suchthat \rdm1 \leq | \Fz_m - \chrg_m | \leq \rdm2 \} ,
       \; \Fz_m \in \mathbb C $.
 The choice of a ring may be useful in a special case when we need
 only the magnitude range and do not need an argument.
 In this case centre of the ring should coincide with the origin of the complex plane
 and we put $ \chrg_m = 0 $.

 Initially we consider the DFT of $ n $-length sequences.
 Enumerative scheme implies recurrent calculation
 of the numbers of these sequences.
 From~\eqref{Eq:Cc} it follows
 that each level of the recursion diminishes $ n $.
 Therefore, we must consider two different types of lengths;
 first is the length of the sequence and is denoted by $ n $;
 second are the lengths of the nested subsequences and are denoted by $ \nr $.
 Moreover, let the spectral properties are given for a large sequence of length $ n $.
 With $ l $ and $ r $ constraints and the shift theorem,
 we can obtain this $ n $-length sequence
 by concatenating $ K \in \mathbb N $ subsequences of length $ \nr $
 such that $ n = K \nr $.

 Also we intend neither to reject nor even to relax the run-length constraints
 for at least two reasons:
 first, these constraints bound the order
 of the recurrence relation~\eqref{Eq:Cc}
 and second, exact values of $ d $ and $ k $ constraints
 should make synchronization control easier.

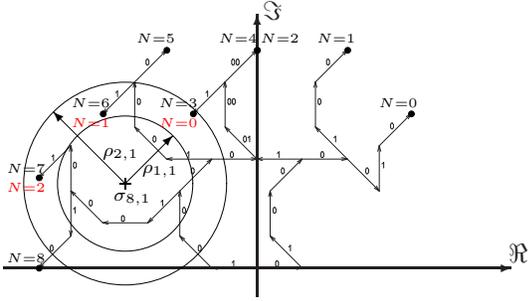
\begin{figure}[!t]
  \begin{center}
  \unitlength =  0.75mm
  \begin{picture}(  90.07 ,  50.07 )(   0.00 ,  40.00 )
   {\thinlines
    \put(  -1.60 ,  45.03 ){ \vector( 1 , 0 ){  90.07 } }
    \put(  43.43 ,  40.00 ){ \vector( 0 , 1 ){  50.07 } } }
    \put(  87.97 ,  46.03 ){ $ \Re $ }
    \put(  44.43 ,  89.07 ){ $ \Im $ }
   \newcommand{\mo}{ \begin{picture}( 1   , 1 )( 0 , -0.07 ) %
                      \drawline( 0.2 , 0.4 )( 0.6 , 0.4 ) 
                      \drawline( 0.7 , 0.7 )( 1   , 1   )( 1   , 0   ) %
                     \end{picture} }
   \newcommand{\po}{ \begin{picture}( 0.5 , 1 )( 0 , -0.07 ) %
                      \drawline( 0.2 , 0.7 )( 0.5 , 1   )( 0.5 , 0   ) %
                     \end{picture} }
   \newcommand{\zr}{ \begin{picture}( 0.8 , 1 )( 0 , -0.07 ) %
                      \drawline( 0.35 , 0    )( 0.2  , 0.25 )( 0.2  , 0.75 ) %
                               ( 0.35 , 1    )( 0.55 , 1    )( 0.7  , 0.75 ) %
                               ( 0.7  , 0.25 )( 0.55 , 0    )( 0.35 , 0    ) %
                     \end{picture} }
   \drawline(  45.03 ,  45.03 )(  53.04 ,  45.03 )
    \drawline(  52.08 ,  45.29 )(  53.04 ,  45.03 )(  52.08 ,  44.78 )
     \put(  48.94 ,  45.23 ){ \llap{ \zr } }
   \drawline(  45.03 ,  45.03 )(  37.03 ,  45.03 )
    \drawline(  37.99 ,  44.78 )(  37.03 ,  45.03 )(  37.99 ,  45.29 )
     \put(  40.93 ,  45.23 ){ \llap{ \po } }
   \drawline(  53.04 ,  45.03 )(  47.38 ,  50.70 )
    \drawline(  47.88 ,  49.83 )(  47.38 ,  50.70 )(  48.25 ,  50.20 )
     \put(  47.15 ,  48.01 ){ \po }
   \drawline(  37.03 ,  45.03 )(  31.37 ,  50.70 )
    \drawline(  31.87 ,  49.83 )(  31.37 ,  50.70 )(  32.23 ,  50.20 )
     \put(  31.14 ,  48.01 ){ \zr }
   \drawline(  47.38 ,  50.70 )(  47.38 ,  58.70 )
    \drawline(  47.12 ,  57.74 )(  47.38 ,  58.70 )(  47.64 ,  57.74 )
     \put(  44.38 ,  54.70 ){ \zr }
   \drawline(  31.37 ,  50.70 )(  31.37 ,  58.70 )
    \drawline(  31.11 ,  57.74 )(  31.37 ,  58.70 )(  31.63 ,  57.74 )
     \put(  28.37 ,  54.70 ){ \zr }
   \drawline(  47.38 ,  58.70 )(  53.04 ,  64.36 )
    \drawline(  52.18 ,  63.86 )(  53.04 ,  64.36 )(  52.54 ,  63.50 )
     \put(  49.97 ,  61.67 ){ \llap{ \zr } }
   \drawline(  31.37 ,  58.70 )(  37.03 ,  64.36 )
    \drawline(  36.16 ,  63.86 )(  37.03 ,  64.36 )(  36.53 ,  63.50 )
     \put(  33.96 ,  61.67 ){ \llap{ \zr } }
   \drawline(  31.37 ,  58.70 )(  25.71 ,  53.04 )
    \drawline(  26.57 ,  53.54 )(  25.71 ,  53.04 )(  26.21 ,  53.91 )
     \put(  28.30 ,  56.01 ){ \llap{ \po } }
   \drawline(  53.04 ,  64.36 )(  61.05 ,  64.36 )
    \drawline(  60.08 ,  64.62 )(  61.05 ,  64.36 )(  60.08 ,  64.10 )
     \put(  56.94 ,  64.56 ){ \llap{ \zr } }
   \drawline(  53.04 ,  64.36 )(  45.03 ,  64.36 )
    \drawline(  46.00 ,  64.10 )(  45.03 ,  64.36 )(  46.00 ,  64.62 )
     \put(  48.94 ,  64.56 ){ \llap{ \po } }
   \drawline(  37.03 ,  64.36 )(  45.03 ,  64.36 )
    \drawline(  44.07 ,  64.62 )(  45.03 ,  64.36 )(  44.07 ,  64.10 )
     \put(  40.93 ,  64.56 ){ \llap{ \zr } }
   \drawline(  37.03 ,  64.36 )(  29.02 ,  64.36 )
    \drawline(  29.99 ,  64.10 )(  29.02 ,  64.36 )(  29.99 ,  64.62 )
     \put(  32.93 ,  64.56 ){ \llap{ \po } }
   \drawline(  25.71 ,  53.04 )(  17.70 ,  53.04 )
    \drawline(  18.67 ,  52.78 )(  17.70 ,  53.04 )(  18.67 ,  53.30 )
     \put(  21.60 ,  53.24 ){ \llap{ \zr } }
   \drawline(  61.05 ,  64.36 )(  66.71 ,  58.70 )
    \drawline(  66.21 ,  59.57 )(  66.71 ,  58.70 )(  65.84 ,  59.20 )
     \put(  60.82 ,  61.67 ){ \zr }
   \drawline(  61.05 ,  64.36 )(  55.39 ,  70.02 )
    \drawline(  55.89 ,  69.16 )(  55.39 ,  70.02 )(  56.25 ,  69.52 )
     \put(  55.16 ,  67.34 ){ \po }
   \drawline(  45.03 ,  64.36 )(  39.37 ,  70.02 )
    \drawline(  39.87 ,  69.16 )(  39.37 ,  70.02 )(  40.24 ,  69.52 )
     \put(  39.15 ,  67.34 ){ \zr }
     \put(  39.95 ,  67.34 ){ \po }
   \drawline(  29.02 ,  64.36 )(  23.36 ,  70.02 )
    \drawline(  23.86 ,  69.16 )(  23.36 ,  70.02 )(  24.23 ,  69.52 )
     \put(  23.13 ,  67.34 ){ \zr }
   \drawline(  17.70 ,  53.04 )(  12.04 ,  58.70 )
    \drawline(  12.54 ,  57.84 )(  12.04 ,  58.70 )(  12.90 ,  58.20 )
     \put(  11.81 ,  56.01 ){ \zr }
   \drawline(  66.71 ,  58.70 )(  66.71 ,  66.71 )
    \drawline(  66.45 ,  65.74 )(  66.71 ,  66.71 )(  66.97 ,  65.74 )
     \put(  63.71 ,  62.71 ){ \po }
   \drawline(  55.39 ,  70.02 )(  55.39 ,  78.03 )
    \drawline(  55.13 ,  77.07 )(  55.39 ,  78.03 )(  55.65 ,  77.07 )
     \put(  52.39 ,  74.03 ){ \zr }
   \drawline(  39.37 ,  70.02 )(  39.37 ,  78.03 )
    \drawline(  39.11 ,  77.07 )(  39.37 ,  78.03 )(  39.63 ,  77.07 )
     \put(  36.37 ,  74.03 ){ \zr }
     \put(  37.17 ,  74.03 ){ \zr }
   \drawline(  23.36 ,  70.02 )(  23.36 ,  78.03 )
    \drawline(  23.10 ,  77.07 )(  23.36 ,  78.03 )(  23.62 ,  77.07 )
     \put(  20.36 ,  74.03 ){ \zr }
   \drawline(  12.04 ,  58.70 )(  12.04 ,  66.71 )
    \drawline(  11.78 ,  65.74 )(  12.04 ,  66.71 )(  12.30 ,  65.74 )
     \put(   9.04 ,  62.71 ){ \zr }
   \drawline(  12.04 ,  58.70 )(  12.04 ,  50.70 )
    \drawline(  12.30 ,  51.66 )(  12.04 ,  50.70 )(  11.78 ,  51.66 )
     \put(   9.04 ,  54.70 ){ \po }
   \drawline(  66.71 ,  66.71 )(  72.37 ,  72.37 )
    \drawline(  71.50 ,  71.87 )(  72.37 ,  72.37 )(  71.87 ,  71.50 )
     \put(  69.30 ,  69.68 ){ \llap{ \zr } }
   \drawline(  55.39 ,  78.03 )(  61.05 ,  83.69 )
    \drawline(  60.18 ,  83.19 )(  61.05 ,  83.69 )(  60.55 ,  82.83 )
     \put(  57.98 ,  81.00 ){ \llap{ \zr } }
   \drawline(  39.37 ,  78.03 )(  45.03 ,  83.69 )
    \drawline(  44.17 ,  83.19 )(  45.03 ,  83.69 )(  44.53 ,  82.83 )
     \put(  41.96 ,  81.00 ){ \llap{ \zr } }
     \put(  41.16 ,  81.00 ){ \llap{ \zr } }
   \drawline(  39.37 ,  78.03 )(  33.71 ,  72.37 )
    \drawline(  34.58 ,  72.87 )(  33.71 ,  72.37 )(  34.21 ,  73.24 )
     \put(  36.30 ,  75.34 ){ \llap{ \po } }
   \drawline(  23.36 ,  78.03 )(  29.02 ,  83.69 )
    \drawline(  28.16 ,  83.19 )(  29.02 ,  83.69 )(  28.52 ,  82.83 )
     \put(  25.95 ,  81.00 ){ \llap{ \zr } }
   \drawline(  23.36 ,  78.03 )(  17.70 ,  72.37 )
    \drawline(  18.57 ,  72.87 )(  17.70 ,  72.37 )(  18.20 ,  73.24 )
     \put(  20.29 ,  75.34 ){ \llap{ \po } }
   \drawline(  12.04 ,  66.71 )(   6.38 ,  61.05 )
    \drawline(   7.24 ,  61.55 )(   6.38 ,  61.05 )(   6.88 ,  61.91 )
     \put(   8.97 ,  64.02 ){ \llap{ \po } }
   \drawline(  12.04 ,  50.70 )(   6.38 ,  45.03 )
    \drawline(   7.24 ,  45.53 )(   6.38 ,  45.03 )(   6.88 ,  45.90 )
     \put(   8.97 ,  48.01 ){ \llap{ \zr } }
   \newcommand{\diam}{ 1.0 }
   \put(  70.77 ,  72.37 ){ \circle*{ \diam } }
   \put(  59.45 ,  83.69 ){ \circle*{ \diam } }
   \put(  43.43 ,  83.69 ){ \circle*{ \diam } }
   \put(  32.11 ,  72.37 ){ \circle*{ \diam } }
   \put(  43.43 ,  83.69 ){ \circle*{ \diam } }
   \put(  27.42 ,  83.69 ){ \circle*{ \diam } }
   \put(  16.10 ,  72.37 ){ \circle*{ \diam } }
   \put(   4.78 ,  61.05 ){ \circle*{ \diam } }
   \put(   4.78 ,  45.03 ){ \circle*{ \diam } }
   \put( 19 , 60 ){ \line( 1 , 0 ){ 2 } }
   \put( 20 , 59 ){ \line( 0 , 1 ){ 2 } }
   \put( 18 , 57 ){ $ \scriptstyle \trig_{ 8 , 1 } $ }
   \put( 20 , 60 ){ \circle{ 24 } }
   \put( 20 , 60 ){ \circle{ 36 } }
   \put( 20 , 60 ){ \vector(  1 , 1 ){ 8.49 } }
   \put( 23 , 62 ){ $ \scriptstyle \rho_{ 1 , 1 } $ }
   \put( 20 , 60 ){ \vector( -1 , 1 ){ 12.73 } }
   \put( 16 , 65 ){ $ \scriptstyle \rho_{ 2 , 1 } $ }
   \put( 65.0 , 73.5 ){ $ \scriptscriptstyle N = 0 $ }
   \put( 54.0 , 85.0 ){ $ \scriptscriptstyle N = 1 $ }
   \put( 44.0 , 85.0 ){ $ \scriptscriptstyle N = 2 $ }
   \put( 26.0 , 73.5 ){ $ \scriptscriptstyle N = 3 $ }
   \put( 36.6 , 85.0 ){ $ \scriptscriptstyle N = 4 $ }
   \put( 22.0 , 85.0 ){ $ \scriptscriptstyle N = 5 $ }
   \put( 10.5 , 73.5 ){ $ \scriptscriptstyle N = 6 $ }
   \put( -1.0 , 62.0 ){ $ \scriptscriptstyle N = 7 $ }
   \put( -1.0 , 46.0 ){ $ \scriptscriptstyle N = 8 $ }
   \put( 26.0 , 70.0 ){ $ \textcolor{red}{\scriptscriptstyle N = 0} $ }
   \put( 10.5 , 70.0 ){ $ \textcolor{red}{\scriptscriptstyle N = 1} $ }
   \put( -1.0 , 58.5 ){ $ \textcolor{red}{\scriptscriptstyle N = 2} $ }
  \end{picture}
  \end{center}
  \vspace{-2ex}
  \caption{Paths and spectral components $ \Fz_1 $ in the complex plane.}
  \label{Fig:Alldklr}
  \vspace{-3ex}
 \end{figure}

 For example, Table~\ref{Tab:Alldklr} shows sequences $\xn$, $\zn$,
 spectral components $\Fz_0$ and $ \Fz_1 $
 respectively.
 The spectral components $ \Fz_1 $ are depicted in Fig.~\ref{Fig:Alldklr}.
 An example of the ring
 $ \mathscr{A}_{ \trig_{ 8 , 1 } , %
                 \rho_{ 1 , 1 } , \rho_{ 2 , 1 } } $
 is superimposed on this image.
 We also show paths that lead from the original to $ \Fz_1 $.
 Assuming that spectrum constraints for $ \Fz_0 $, $ \Fz_2 \dots \Fz_7 $ are relaxed,
 we summarize in Table~\ref{Tab:FSdklr} those sequences
 for which components $ \Fz_1 $ lie in the ring
 $ \mathscr{A}_{ \trig_{ 8 , 1 } , %
                 \rho_{ 1 , 1 } , \rho_{ 2 , 1 } } $.

 \begin{table}[!t]
   \centering
  \begin{threeparttable}
   \caption{All Lexicographically Ordered $dklr$ Sequences
    of Length $n=8$. Constraints: $d=2$, $k=4$, $l=1$, $r=3$.}
   \label{Tab:Alldklr}
   \tabcolsep=0.5em
   \renewcommand{\arraystretch}{0}
   \newcommand{\z}{$\scriptstyle1$}
   \newcommand{\xs}{\rule{0pt}{2ex}}
   \newcommand{\zs}{\rule{0pt}{1ex}}
   \newcommand{\zf}{\rule{0pt}{2pt}}
   \begin{tabular}{|c|cccccccc|c|c|}
    \hline
    \strut %
    $N\tnote{a}$ %
        & $x_1$ & $x_2$ & $x_3$ & $x_4$ & %
          $x_5$ & $x_6$ & $x_7$ & $x_8$ & & \\
        & $z_1$ & $z_2$ & $z_3$ & $z_4$ & %
          $z_5$ & $z_6$ & $z_7$ & $z_8$ & $ \Fz_0 $ & $ \Fz_1 $ \\
    \hhline{|=|========|=|=|}
    \xs%
    0  &  0&  1&  0&  0&  0&  0&  1&  0 &    &               \\
    \zs%
       & \z&-\z&-\z&-\z&-\z&-\z& \z& \z & -2 & ( 3.41, 3.41) \\
    \zf&   &   &   &   &   &   &   &    &    &               \\
    \hline
    \xs%
    1  &  0&  1&  0&  0&  0&  1&  0&  0 &    &               \\
    \zs%
       & \z&-\z&-\z&-\z&-\z& \z& \z& \z &  0 & ( 2.00, 4.83) \\
    \zf&   &   &   &   &   &   &   &    &    &               \\
    \hline
    \xs%
    2  &  0&  1&  0&  0&  1&  0&  0&  0 &    &               \\
    \zs%
       & \z&-\z&-\z&-\z& \z& \z& \z& \z &  2 & ( 0.00, 4.83) \\
    \zf&   &   &   &   &   &   &   &    &    &               \\
    \hline
    \xs%
    3  &  0&  1&  0&  0&  1&  0&  0&  1 &    &               \\
    \zs%
       & \z&-\z&-\z&-\z& \z& \z& \z&-\z &  0 & (-1.41, 3.41) \\
    \zf&   &   &   &   &   &   &   &    &    &               \\
    \hline
    \xs%
    4  &  1&  0&  0&  0&  0&  1&  0&  0 &    &               \\
    \zs%
       &-\z&-\z&-\z&-\z&-\z& \z& \z& \z & -2 & ( 0.00, 4.83) \\
    \zf&   &   &   &   &   &   &   &    &    &               \\
    \hline
    \xs%
    5  &  1&  0&  0&  0&  1&  0&  0&  0 &    &               \\
    \zs%
       &-\z&-\z&-\z&-\z& \z& \z& \z& \z &  0 & (-2.00, 4.83) \\
    \zf&   &   &   &   &   &   &   &    &    &               \\
    \hline
    \xs%
    6  &  1&  0&  0&  0&  1&  0&  0&  1 &    &               \\
    \zs%
       &-\z&-\z&-\z&-\z& \z& \z& \z&-\z & -2 & (-3.41, 3.41) \\
    \zf&   &   &   &   &   &   &   &    &    &               \\
    \hline
    \xs%
    7  &  1&  0&  0&  1&  0&  0&  0&  1 &    &               \\
    \zs%
       &-\z&-\z&-\z& \z& \z& \z& \z&-\z &  0 & (-4.83, 2.00) \\
    \zf&   &   &   &   &   &   &   &    &    &               \\
    \hline
    \xs%
    8  &  1&  0&  0&  1&  0&  0&  1&  0 &    &               \\
    \zs%
       &-\z&-\z&-\z& \z& \z& \z&-\z&-\z & -2 & (-4.83, 0.00) \\
    \zf&   &   &   &   &   &   &   &    &    &               \\
    \hline
   \end{tabular}
   \begin{tablenotes}
    \item [a] By $N$ we denote the lexicographic index of sequence.
   \end{tablenotes}
  \end{threeparttable}
 \end{table}

 \begin{table}[!t]
   \centering
  \begin{threeparttable}
   \caption{All Lexicographically Ordered Spectrum Shaped Sequences
    from Table~\ref{Tab:Alldklr} With
    $ \trig_{ 8 , 1 } = ( -2.93 , 1.87 ) $,
    $ \rho_{ 1 , 1 } = 1.5 $, $ \rho_{ 2 , 1 } = 2.25 $.}
   \label{Tab:FSdklr}
   \tabcolsep=0.5em
   \begin{tabular}{|r|cccccccc|c|}
    \hline
    $ \textcolor{red}{N}^{\vphantom{X}} $ %
        & $x_1$ & $x_2$ & $x_3$ & $x_4$ & %
          $x_5$ & $x_6$ & $x_7$ & $x_8$ & $ \Fz_1 $ \\
    \hline
    \rule{0pt}{2ex}
    \textcolor{red}{0} & 0 & 1 & 0 & 0 & 1 & 0 & 0 & 1 & (-1.41, 3.41) \\
    \textcolor{red}{1} & 1 & 0 & 0 & 0 & 1 & 0 & 0 & 1 & (-3.41, 3.41) \\
    \textcolor{red}{2} & 1 & 0 & 0 & 1 & 0 & 0 & 0 & 1 & (-4.83, 2.00) \\
    \hline
   \end{tabular}
  \end{threeparttable}
 \vspace{-5ex}
 \end{table}

 \section{The Number of Sequences} \label{Sec:Recur}

 By $ \prou = e^{ \frac{ 2 \pi i } n } $
 denote a primitive $ n $th root of unity.
 Now we shall give the following definition.
 Let $ 0 \leq m \leq n - 1 $,
   $ \nr \leq n $,
   $ z \in \{ - 1 , 1 \} $,
 and let $ \phase \in \mathbb C $ be some constant.
 Then an order $ j $ recurrence relation
 \[
  \trigntm = z \sum_{ t = 0 }^{ j - 1 }
                \prou^{ - m t }
               + \phase \trigntjm
 \]
 is called a \emph{trigonometric recurrence relation}.
 The constant $ \phase $ is said to be a linear phase if
 $ \phase = z \prou^{ - m j } $.

 This trigonometric recurrence relation is not basically very different
 from the Danielson-Lanczos identity~\cite{DanielsonLanczos42}.
 There, an $ n $-point DFT was split into the sum of two DFT;
 one is formed from the even-numbered points,
 the other from the odd-numbered points.
 Here, we add an exponential sum of the prefix
 to multiplying by the linear phase
 trigonometric recurrence relation of the subsequence.

 Consider bipolar run-length constrained sequences of length $ \nr $ and having
 an $ m $th trigonometric recurrence relation ring $ \annn $
 centered at $ \trigntm $ with inner radius $ \rdm1 $
 and outer radius $ \rdm2 $; here the ring
 $ \annn $ is a set
 $  \annn
   = \{ \trigntm
      \suchthat \rdm1 \leq | \Fz_m - \trigntm | \leq \rdm2 \} ,
       \; \Fz_m \in \mathbb C $.

 We compute the number of $ \nr $-length subsequences of $ \zn $.

 Let  $ \FSS_{ n , \nr }^{ \trigntm , \rdm1 , \rdm2 } $
 be the number of these subsequences, which begin with one.
 Let $ \FSSl_{ n , \nr }^{ \trigntm , \rdm1 , \rdm2 } $
 be the number of these sequences, which begin with a leading run of zeros.

 Since an internal run of zeros succeeds a leading run of zeros,
 we see that the leading constraint $l$ does not affect
 $  \FSS_{ n , \nr , m }^{ \trigntm , \rdm1 , \rdm2 } $.
 For convenience, below, under
 $  \FSS_{ n , \nr , m }^{ \trigntm , \rdm1 , \rdm2 } $
 we imply
 $  \FSS_{ n , \nr , m }^{ \trigntm , \rdm1 , \rdm2 } (d, k, r) $
 and under
 $ \FSSl_{ n , \nr , m }^{ \trigntm , \rdm1 , \rdm2 } $
 we similarly imply
 $ \FSSl_{ n , \nr , m }^{ \trigntm , \rdm1 , \rdm2 } (d, k, l, r) $.

 \begin{proposition}
  \label{Prop:C}
  The numbers $ \FSS_{ n , \nr , m }^{ \trigntm , \rdm1 , \rdm2 } $
  and $ \FSSl_{ n , \nr , m }^{ \trigntm , \rdm1 , \rdm2 } $
  can be obtained as:\\
  \begin{multline}
   \label{Eq:C_recur}
   \FSS_{ n , \nr , m }^{ \trigntm , \rdm1 , \rdm2 } \\
    = \begin{cases}
         a_{ \nr , m } ,                            & \nr < d + 1 , \\
         \sum_{ j = d + 1 }^{ \min( \nr , k + 1 ) }
          \FSS_{ n , \nr - j , m }^{ \trigntjm ,
                                 \rdm1 , \rdm2 }
       + b_{ \nr , m } ,                            &       d + 1 \leq \nr ,
      \end{cases}
  \end{multline}
  where
  \begin{equation}
   \label{Eq:trig_n_j}
   \trigntjm
    =             - \prou^{   m j }
      \left(
                    \trigntm
       + \prou^{ - m ( j - 1 ) / 2 }
         \frac{ \sin \left( \frac{ m j } n \pi \right) }
              { \sin \left( \frac  m     n \pi \right) }
      \right) ,
  \end{equation}
  initial condition
  \begin{equation}
   \label{Eq:C_a}
   a_{ \nr , m } = \begin{cases}
                     1 , & \nr \leq r + 1 %
                            \text{\textnormal{ and }} %
                           \rdm1 \leq \left|
                                       \trigntm - \trigntmR
                                      \right| \leq \rdm2 , \\
                     0 , & \text{otherwise} ,
                   \end{cases}
  \end{equation}
  and additional condition
  \begin{equation}
   \label{Eq:C_b}
   b_{ \nr , m } = \begin{cases}
                    - 1 , & \begin{gathered}[t]
                             r + 1 < \nr \leq k + 1 \\
                              \text{\textnormal{ and }} %
                             \rdm1 \leq \left|
                                         \trigntm - \trigntmR
                                        \right| \leq \rdm2 ,
                            \end{gathered} \\ 
                      1 , & \begin{gathered}[t]
                             k + 1 < \nr \leq r + 1 \\
                              \text{\textnormal{ and }} %
                             \rdm1 \leq \left|
                                         \trigntm - \trigntmR
                                        \right| \leq \rdm2 ,
                            \end{gathered} \\ 
                      0 , & \text{otherwise} ,
                   \end{cases}
  \end{equation}
  where
  \begin{equation}
   \label{Eq:Dirichlet_kernel}
   \trigntmR
      = - \prou^{ - m ( \nr - 1 ) / 2 }
           \frac{ \sin \left( \frac{ m \nr } n \pi \right) }
                { \sin \left( \frac  m       n \pi \right) } .
  \end{equation}

  If a leading series is running, then
  \begin{equation*}
   \FSSl_{ n , \nr , m }^{ \trigntm , \rdm1 , \rdm2 }
    =    \sum_{ j = 0 }^{ \min( \nr , l ) }
          \FSS_{ n , \nr - j }^{ \trigntmL ,
                                 \rdm1 , \rdm2 }
       + \hat b_{ \nr , m } ,
  \end{equation*}
  where
  \begin{equation}
   \label{Eq:LR_trig_n_j}
   \trigntmL
    =               \prou^{   m j }
      \left(
                    \trigntm
       - \prou^{ - m ( j - 1 ) / 2 }
         \frac{ \sin \left( \frac{ m j } n \pi \right) }
              { \sin \left( \frac  m     n \pi \right) }
      \right) ,
  \end{equation}
  additional condition
  \begin{equation}
   \label{Eq:Cl_b}
   \hat b_{ \nr , m } = \begin{cases}
                         - 1 , & \begin{gathered}[t]
                                  r < \nr \leq l %
                                   \text{\textnormal{ and }} \\
                                  \rdm1 \leq \left|
                                              \trigntm + \trigntmR
                                             \right| \leq \rdm2 ,
                                 \end{gathered} \\ 
                           0 , & \text{otherwise} .
                        \end{cases}
  \end{equation}
 \end{proposition}
 Here $ d \geq 0 $, $ k \geq d $, $ l \geq 0 $, $ r \geq 0 $.
 \begin{proof}
  First, consider initial and additional conditions~\eqref{Eq:C_a}
                                                and~\eqref{Eq:C_b}.
  In the case of $ \nr \leq r + 1 $,
  there is only a trailing run in the subsequence.
  This trailing subsequence of $ \xn $ consists of one and
  $ \nr - 1 $ zeros.
  The NRZI rule takes each term of the trailing subsequence to $ - 1 $.
  It defines the $ m $th exponential sum $ \trigntmR $,
  which corresponds to this subsequence, as follows:
  \[
   \newcommand{ \mz }{ \scriptstyle{ - 1 } }
   \underbrace{ \overbrace{ \begin{array}{@{}r@{\,}r@{\,}r@{\,}r@{}}
                               1 &   0 & \dots &   0 \\
                             \mz & \mz & \dots & \mz
                            \end{array}
                          }^{ 1 \leq \nr \leq r + 1 }
              }_{ \trigntmR = \sum_{ t = 0 }^{ \nr - 1 }
                               ( - 1 ) \prou^{ - m t } %
                                        \displaystyle . }
  \]
  Therefore, it gives us the only allowed sequence
  which length lies in the interval $ [ 1 , r + 1 ] $
  and $ \trigntmR $ lies in the ring $ \annn $, i.e.
  \begin{equation}
   \label{Eq:in_ring}
   \rdm1 \leq \left| \trigntm - \trigntmR \right| \leq \rdm2 .
  \end{equation}
  Consider a finite geometric series;
  then recall the derivation of the Dirichlet kernel trigonometric identity
  and obtain
  \begin{align}
   \label{Eq:GeomSer}
   \sum_{ t = 0 }^{ \nr - 1 }
    \prou^{ - m t }
    & = \frac{ 1 - \prou^{ - m \nr } }
             { 1 - \prou^{ - m   } } \\
   \label{Eq:Dirichlet_identity}
    & = \prou^{ - m ( \nr - 1 ) / 2 }
         \frac{ \sin \left( \frac{ m \nr } n \pi \right) }
              { \sin \left( \frac  m       n \pi \right) } .
  \end{align}
  Multiplying~\eqref{Eq:Dirichlet_identity} by $ - 1 $,
  we obtain~\eqref{Eq:Dirichlet_kernel}.

  Now we must prove the recurrence relation in~\eqref{Eq:C_recur}.
  According to Cover's enumerative method~\cite{Cover73},
  we build the recursion by the following way.
  Let us consider a possible run of zeros, which follows the leading one,
  as a prefix for the following subsequences beginning also with one.
  Assuming the length $ j $ of the prefix
  grows from $ d + 1 $ to $ \min( \nr , k + 1 ) $
  and weight of this prefix equals one,
  we can consider a concatenation of the prefix
  and the following subsequences as
  \[
   \newcommand{ \mz }{ \scriptstyle{ - 1 } }
   \newcommand{ \pz }{ \scriptstyle{ \hphantom{ - } 1 } }
   \underbrace{ \overbrace{ \begin{array}{@{}r@{\,}r@{\,}r@{\,}r@{}}
                               1 &   0 & \dots &   0 \\
                             \mz & \mz & \dots & \mz
                            \end{array}
                          }^{ d + 1 \leq j \leq k + 1 }
               \underbrace{ \begin{array}{@{}r@{\,}r@{\,}r@{}}
                               1 &   0 & \dots \\
                             \pz & \pz & \dots
                            \end{array}
                          }_{ \trigntjm }
              }_{ \trigntm =   \sum_{ t = 0 }^{ j - 1 }
                                ( - 1 ) \prou^{ - m t }
                               + \phase \trigntjm %
                                        \displaystyle . }
  \]
  In fact, if the subsequences of length $ \nr $ begin with one,
  then the $ m $th trigonometric recurrence relation for $ \trigntm $
  can be obtained as
  \[
   \trigntm = - \sum_{ t = 0 }^{ j - 1 }
                 \prou^{ - m t }
                + \phase \trigntjm ,
  \]
  where the first term is the $ m $th exponential sum of the prefix and
  $ \trigntjm $ is the $ m $th trigonometric recurrence relation
  corresponding to following subsequence;
  this subsequence also begins with one.
  The phase factor $ \phase $ can be found by the shift theorem as follows:
  Before concatenating a prefix,
  for the first term $ z_1 $ of $ \zn $, we have
  \[
   z_1 \prou^{ - m 0 } = z_1 .
  \]
  Since there is just a one in the prefix,
  it follows that this prefix changes a sign of each term in the following subsequence.
  Therefore, after concatenating a prefix of length $ j $,
  for the same term we obtain
  \[
   \phase z_1 = - z_{ j + 1 } \prou^{ - m j } .
  \]
  Since   $ z_1 $ corresponds to $ x_1 $ and
  $ z_{ j + 1 } $ corresponds to $ x_{ j + 1 } $,
  where $ x_1 $ and $ x_{ j + 1 } $ are the same term, it follows that
  \[
   \phase = - \prou^{ - m j } .
  \]
  Therefore
  \begin{equation}
   \label{Eq:trig_recur}
   \trigntm = - \sum_{ t = 0 }^{ j - 1 }
                 \prou^{ - m t }
              - \prou^{ - m j }
                 \trigntjm ,
  \end{equation}
  Consider a finite geometric series in this equation; then
  \begin{align*}
   \sum_{ t = 0 }^{ j - 1 }
    \prou^{ - m t }
    & = \frac{ 1 - \prou^{ - m j } }
             { 1 - \prou^{ - m   } } \\
    & = \prou^{ - m ( j - 1 ) / 2 }
         \frac{ \sin \left( \frac{ m j } n \pi \right) }
              { \sin \left( \frac  m     n \pi \right) } ,
  \end{align*}
  and substituting it for the sum in~\eqref{Eq:trig_recur}, we get
  \[
     \prou^{ - m j } \trigntjm
      = - \trig_{ \nr }
        - \prou^{ - m ( j - 1 ) / 2 }
           \frac{ \sin \left( \frac{ m j } n \pi \right) }
                { \sin \left( \frac  m     n \pi \right) } .
  \]
  Multiplying both sides by $ \prou^{   m j } $,
  we obtain~\eqref{Eq:trig_n_j}.

  The other case is when a leading series of zeros is running.
  First, we also consider the additional condition.
  In the case of $ \nr \leq \min( l , r ) $,
  the leading run of zeros is the trailing one.
  This also gives us the only allowed sequence
  which length lies in the interval $[0, \min(l, r)]$.
  \[
   \newcommand{ \pz }{ \scriptstyle{ 1 } }
   \underbrace{ \overbrace{ \begin{array}{@{}c@{\,}c@{\,}c@{}}
                               0 & \dots &   0 \\
                             \pz & \dots & \pz
                            \end{array}
                          }^{ \lefteqn{ \scriptstyle
                              0 \leq \nr \leq \min( l , r ) } }
              }_{ \trigntm   = \sum_{ t = 0 }^{ \nr - 1 }
                                \prou^{ - m t } %
                                        \displaystyle . }
  \]
  Using~\eqref{Eq:GeomSer} we get
  \begin{align*}
   \trig_{ \nr }
    & = \frac{ 1 - \prou^{ - m \nr } }
             { 1 - \prou^{ - m   } } \\
    & = \prou^{ - m ( \nr - 1 ) / 2 }
         \frac{ \sin \left( \frac{ m \nr } n \pi \right) }
              { \sin \left( \frac  m       n \pi \right) } .
  \end{align*}
  Substituting it for $ \trigntm $ in~\eqref{Eq:in_ring},
  we obtain the condition in~\eqref{Eq:Cl_b}.

  In the case of nonzero weight, there exist only zero weight prefixes
  which length lies in the interval $ [ 0 , l ] $.
  Subsequences beginning with one follow the prefixes;
  therefore, we can consider a concatenation of this prefix
  and the following subsequences as
  \[
   \newcommand{ \mz }{ \scriptstyle{ - 1 } }
   \newcommand{ \pz }{ \scriptstyle{ 1 } }
   \underbrace{ \overbrace{ \begin{array}{@{}c@{\,}c@{\,}c@{}}
                               0 & \dots &   0 \\
                             \pz & \dots & \pz
                            \end{array}
                          }^{ 0 \leq j \leq l }
               \underbrace{ \begin{array}{@{}r@{\,}r@{\,}r@{}}
                               1 &   0 & \dots \\
                             \mz & \mz & \dots
                            \end{array}
                          }_{ \trigntmL }
              }_{ \trigntm   =   \sum_{ t = 0 }^{ j - 1 }
                                  \prou^{ - m t }
                                 + \phase \trigntmL %
                                        \displaystyle . }
  \]
  Thus, the $ m $th expression for $ \trigntm $ is
  \[
   \trigntm =   \sum_{ t = 0 }^{ j - 1 }
                 \prou^{ - m t }
                + \phase \trigntmL .
  \]
  Recall the reasoning that led us to~\eqref{Eq:trig_n_j}.
  Arguing as there,
  and taking into account that the prefix now consists of all zeros,
  we get $ \phase = \prou^{ - m j } $ and
  \[
   - \prou^{ - m j } \trigntmL
      = - \trigntm
        - \prou^{ - m ( j - 1 ) / 2 }
           \frac{ \sin \left( \frac{ m j } n \pi \right) }
                { \sin \left( \frac  m     n \pi \right) } .
  \]
  Multiplying both sides by $ - \prou^{   m j } $,
  we obtain~\eqref{Eq:LR_trig_n_j}.
 \end{proof}

 \newcommand{\trign}{\boldsymbol{\trig}}
 \newcommand{\innrn}{\boldsymbol{\rho}_1}
 \newcommand{\outrn}{\boldsymbol{\rho}_2}
 Now consider a set $ \annS $ of the rings
 $ \annj0 , \annj1 , \dots , \annj{ n - 1 } $.
 Let $ \trign_{ \nr } = ( \trig_{ \nr , 0 } ,
                          \trig_{ \nr , 1 } ,
                                      \dots ,
                          \trig_{ \nr , n - 1 } ) %
       \in \mathbb C^n $
 be a vector of centres of these rings,
 $ \innrn = ( \rho_{ 1 , 0 } ,
              \rho_{ 1 , 1 } ,
                       \dots ,
              \rho_{ 1 , n - 1 } ) %
   \in \mathbb R^n $ and
 $ \outrn = ( \rho_{ 2 , 0 } ,
              \rho_{ 2 , 1 } ,
                       \dots , 
              \rho_{ 2 , n - 1 } ) %
   \in \mathbb R^n $ vectors of theirs inner and outer radii.

 Then consider those $ \nr $-length subsequences from $ \Sdkr( n ) $
 whose vectors of the exponential sums belong to $ \annS $.

 Let  $ \FSS_{ n , \nr }^{ \trign_{ \nr } , \innrn , \outrn } $
 and $ \FSSl_{ n , \nr }^{ \trign_{ \nr } , \innrn , \outrn } $
 be the number of these sequences,
 which begin with one and with a leading run of zeros respectively.

 As above, under
 $  \FSS_{ n , \nr }^{ \trign_{ \nr } , \innrn , \outrn } $
 we imply
 $  \FSS_{ n , \nr }^{ \trign_{ \nr } , \innrn , \outrn } (d, k, r) $
 and under
 $ \FSSl_{ n , \nr }^{ \trign_{ \nr } , \innrn , \outrn } $
 we similarly imply
 $ \FSSl_{ n , \nr }^{ \trign_{ \nr } , \innrn , \outrn } (d, k, l, r) $.

 We state without proof the following proposition.
 \begin{proposition}
  \label{Prop:Cvect}
  The numbers $ \FSS_{ n , \nr }^{ \trign_{ \nr } , \innrn , \outrn } $
  and        $ \FSSl_{ n , \nr }^{ \trign_{ \nr } , \innrn , \outrn } $
  can be expressed as:\\
  \begin{equation}
   \label{Eq:Cvect_recur}
   \FSS_{ n , \nr }^{ \trign_{ \nr } , \innrn , \outrn } %
    = \begin{cases}
         a_{ \nr } ,                                & \nr < d + 1 , \\
         \sum_{ j = d + 1 }^{ \min( \nr , k + 1 ) }
          \FSS_{ n , \nr - j }^{ \trign_{ \nr - j } ,
                                 \innrn , \outrn }
       + b_{ \nr } ,                                &       d + 1 \leq \nr ,
      \end{cases}
  \end{equation}
  where components $ \trig_{ \nr - j , m } $
  of the vector $ \trign_{ \nr - j } = ( \trig_{ \nr - j , 0 } ,
                                         \trig_{ \nr - j , 1 } ,
                                                         \dots ,
                                         \trig_{ \nr - j , n - 1 } ) $
  are defined 
  by~\eqref{Eq:trig_n_j}.
  Here initial condition
  \begin{equation}
   \label{Eq:Cvect_a}
   a_{ \nr } = \begin{cases}
                 1 , & \nr \leq r + 1 %
                        \text{\textnormal{ and }} %
                        \inring \text{ is true} , \\
                 0 , & \text{otherwise} ,
               \end{cases}
  \end{equation}
  and additional condition
  \begin{equation}
   \label{Eq:Cvect_b}
   b_{ \nr } = \begin{cases}
                - 1 , & r + 1 < \nr \leq k + 1 %
                         \text{\textnormal{ and }} %
                        \inring \text{ is true} , \\
                  1 , & k + 1 < \nr \leq r + 1 %
                         \text{\textnormal{ and }} %
                        \inring \text{ is true} , \\
                  0 , & \text{otherwise} ,
               \end{cases}
  \end{equation}
  where the indicator function $ \inring $ is the logical conjunction
  of $ n $ statements each of which predicates of the
  $ ( m ) $th exponential sum of trailing run~\eqref{Eq:Dirichlet_kernel}
  lies in the ring $ \annn $, i.e.
  \[
   \inring = \bigwedge_{ m = 0 }^{ n - 1 }
              \left(
               \rho_{ 1 , m } \leq \left|
                                    \trig_{ \nr , m } - \trigR_{ \nr , m }
                                   \right| \leq \rho_{ 2 , m }
                                   \mathstrut^{ \mathstrut }
              \right) .
  \]

  If a leading series is running, then
  \begin{equation}
   \label{Eq:Cvectl}
   \FSSl_{ n , \nr }^{ \trign_{ \nr } , \innrn , \outrn }
    =    \sum_{ j = 0 }^{ \min( \nr , l ) }
          \FSS_{ n , \nr - j }^{ \hat \trign_{ \nr - j } ,
                                 \innrn , \outrn }
       + \hat b_{ \nr } ,
  \end{equation}
  where components $ \hat \trig_{ \nr - j , m } $
  of the vector $ \hat \trign_{ \nr - j } = ( \hat \trig_{ \nr - j , 0 } ,
                                              \hat \trig_{ \nr - j , 1 } ,
                                                                   \dots ,
                                              \hat \trig_{ \nr - j , n - 1 } ) $
  are defined 
  by~\eqref{Eq:LR_trig_n_j}.
  Here additional condition
  \begin{equation*}
   \label{Eq:Cvectl_b}
   \hat %
   b_{ \nr } = \begin{cases}
                - 1 , & r < \nr \leq l %
                         \text{\textnormal{ and }} %
                        \hat %
                        \inring \text{ is true} , \\
                  0 , & \text{otherwise} ,
               \end{cases}
  \end{equation*}
  where the indicator function $ \hat \inring $ is the logical conjunction
  of $ n $ statements each of which predicates of the
  $ ( m ) $th exponential sum of trailing run
  $ - \trigR_{ \nr , m } $
  lies in the ring $ \annn $, i.e.
  \[
   \hat %
   \inring = \bigwedge_{ m = 0 }^{ n - 1 }
              \left(
               \rho_{ 1 , m } \leq \left|
                                    \trig_{ \nr , m } + \trigR_{ \nr , m }
                                   \right| \leq \rho_{ 2 , m }
                                   \mathstrut^{ \mathstrut }
              \right) .
  \]
 \end{proposition}

 \section{Algorithms for Encoding and Decoding
  Spectrum Shaped
  $ dklr $ Sequences}
  \label{Sec:EnumAlg}

 Now let $\Sdklr$ denotes a set of spectrum shaped
 binary run-length constrained sequences
 $\xn=(x_1, x_2, \dots,$ $ x_{ \nr } )$ of length $ \nr $.
 Let the set $\Sdklr$ be ordered lexicographically.
 From it follows that the lexicographic index
 $ N( \xn ) \in \{ \mathbb N , 0 \} $ of
 $\xn \in \Sdklr$ is given by
 \begin{equation*}
  N(\xn)=\sum_{j=1}^{ \nr } x_j W ( \pn ) ,
 \end{equation*}
 where $W ( \pn )$ denotes the number of sequences
 in $\Sdklr$ with given prefix $ \pn = (x_1, x_2, \dots, x_{j-1}, 0) $.

 The decoding algorithm, for given sequence $\xn$,
 find its lexicographic index $ N( \xn ) < |\Sdklr|$.
 This is done by successive approximation method
 using $W ( \pn )$ as the weight of term $ x_j $.

 By $ a_j ( \pn ) $ denote the number of trailing zeros of the prefix \pn.
 By $ \nu_{ j - 1 } = \sum_{ i = 1 }^{ j - 1 } x_i $
 denote the weight of this prefix.
 Since $ \pn $ is the prefix of $ \xn $,
 it follows that subsequence $ \tilde \xn = (x_j, x_{j+1}, \dots, x_{ \nr } ) $
 is the rest of $ \xn $,
 and $ l_j $ is the leading run of zeros in this subsequence.
 We define $ l_j $ as the complement of $ a_j ( \pn ) $
 in $ l $ (for a leading run  of zeros) or
 in $ k $ (for the other runs of zeros) as follows:
 \[
  l_j=\begin{cases}
   l-a_j(\pn), & \nu_{ j - 1 } = 0 , \\
   k-a_j(\pn), & \text{otherwise} .
  \end{cases}
 \]

 Now we can compute the number of the spectrum shaped sequences
 $ W_{ \trign_{ \nr } } ( \pn ) $ as
 \[
  W ( \pn )
  = \begin{cases}
       \FSSl_{ n , \nr -j}^{\trign_{ \nr -j} , %
                            \innrn , \outrn }(d, k, l_j, r )
     - \tilde b_{ \nr },                                     & l_j \geq 0 ,    \\
     0,                                                      & \text{otherwise} .
    \end{cases}
 \]
 By $ \tilde b_{ \nr } $ we take into account a trailing run as follows:
 \[
  \tilde %
  b_{ \nr } = \begin{cases}
               1 , & r - a_j  ( \pn ) < \nr - j \leq \min( l_j , r )
                      \text{ and } %
                     \tilde %
                     \inring \text{ is true} , \\
               0 , & \text{otherwise} ,
              \end{cases}
 \]
 where the indicator function $ \tilde \inring $ is the logical conjunction
 of $ n $ statements each of which predicates of
 the $ m $th exponential sum of the trailing run
 \[
  \trigR_{ \nr - j , m }
     =   \prou^{ - m ( \nr - j - 1 ) / 2 }
          \frac{ \sin \left( \frac{ m ( \nr - j ) } n \pi \right) }
               { \sin \left( \frac  m               n \pi \right) }
 \]
 lies in the ring $ \annnj $, i.e.
 \[
  \tilde %
  \inring = \bigwedge_{ m = 0 }^{ n - 1 }
             \left(
              \rho_{ 1 , m } \leq \left|
                                   \trig_{ \nr - j , m } - \trigR_{ \nr - j , m }
                                  \right| \leq \rho_{ 2 , m }
                                  \mathstrut^{ \mathstrut }
             \right) ,
 \]
 where components $ \trig_{ \nr - j , m } $
 of the vector $ \trign_{ \nr - j } = ( \trig_{ \nr - j , 0 } ,
                                        \trig_{ \nr - j , 1 } ,
                                                        \dots ,
                                        \trig_{ \nr - j , n - 1 } ) $
  are defined as
 \[
  \trig_{ \nr - j , m }
   =   ( - 1 )^{ \nu_{ j - 1 } } \prou^{ m j }(   \trig_{ \nr , m }
                                                - \trig_{ j - 1 , m } )
     - \prou^m .
 \]

 \newcommand{\trigp}{\boldsymbol{c}}
 We introduce the next variables: $a$, $w$, $ \trigp = (c_0, c_1, \dots,$ $c_{n-1} )$,
 which correspond to $ a_j  ( \pn ) $, $ \nu_{ j - 1 } $, $ \trign_{ j - 1 } $.

 \begin{tabbing}
  \quad \= \quad \= \quad \= \+ \kill
  $ N:=0; \quad a:=1; \quad w:=0; \quad \trigp := 0 ; $ \\*
  \textbf{for} $j:=1$ \textbf{to} $ \nr $ \textbf{do} \+ \\*
    \pushtabs
    Get $W ( \pn );$\\*
    \poptabs
    \textbf{if} $ x_j = 1 $ \textbf{then}\\*
      \> $N:=N+W ( \pn ); \quad a:=1; \quad w:=w+1;$\\*
    \textbf{else}\\*
      \> $a:=a+1;$\\*
    \textbf{end if \dots else} \\*
    \textbf{for} $ m := 0 $ \textbf{to} $ n - 1 $ \textbf{do} \+ \\*
      $ c_m := c_m + ( -1 )^w \prou^{ - m ( j - 1 ) } ; $ \- \\*
    \textbf{end for} \- \\*
  \textbf{end for}.
 \end{tabbing}

 The encoding (inverse) algorithm, for given lexicographic index
 $ N( \xn ) < |\Sdklr|$, find the corresponding $\xn$.
 \begin{tabbing}
  \quad \= \quad \= \quad \= \+ \kill
  $ a:=1; \quad w:=0; \quad \trigp := 0 ; $ \\*
  \textbf{for} $j:=1$ \textbf{to} $ \nr $ \textbf{do} \+ \\*
    \pushtabs
    Get $W ( \pn );$\\*
    \poptabs
    \textbf{if} $ N \geq W ( \pn ) $ \textbf{then}\\*
      \> $N:=N-W ( \pn ); \quad x_j:=1; \quad a:=1; \quad w:=w+1;$\\*
    \textbf{else}\\*
      \> $x_j:=0; \quad a:=a+1;$\\*
    \textbf{end if \dots else}\\*
    \textbf{for} $ m := 0 $ \textbf{to} $ n - 1 $ \textbf{do} \+ \\*
      $ c_m := c_m + ( -1 )^w \prou^{ - m ( j - 1 ) } ; $ \- \\*
    \textbf{end for} \- \\*
   \textbf{end for}.
 \end{tabbing}

 \section{Conclusion}

 We presented an enumerative approach for constructing binary
 run-length limited sequences to satisfy spectral constraints.
 By considering the spectral components of the DFT in the complex plane,
 we got the possibility to use Cover's enumerative scheme for encoding
 these spectrum shaped sequences.
 First, we defined the sequences whose $ m $th spectral components
 lie in a certain ring.
 Secondly, we proposed recurrence relations for calculating
 the number of such sequences.
 Then, we expanded our method for vectors of all spectral components.
 Finally, we suggested algorithms for encoding and decoding these sequences.

 \bibliographystyle{IEEEtran}
 \bibliography{rll-seq}
\end{document}